\documentclass[twocolumn]{aastex631}

\received{September 9, 2024}
\accepted{\today}

\submitjournal{ApJ}

\begin{document}

\title{Sigmoid eruption associated with X9.3 flare from AR 12673 drives gradual SEP event on 2017 September 6}


\correspondingauthor{Stephanie L. Yardley}
\email{steph.yardley@northumbria.ac.uk}

\author[0000-0002-0786-7307]{Stephanie L. Yardley}
\affiliation{Department of Mathematics, Physics and Electrical
Engineering, Northumbria University, Ellison Place, Newcastle
Upon Tyne, NE1 8ST, UK}
\affiliation{University College London, Mullard Space Science Laboratory, Holmbury St. Mary, Dorking, Surrey, RH5 6NT, UK}
\affiliation{Donostia International Physics Center (DIPC), Paseo Manuel de Lardizabal 4, 20018 San Sebasti{\'a}n, Spain}

\author[0000-0002-2189-9313]{David H. Brooks}
\affiliation{Computational Physics Inc., Springfield, VA 22151, USA}

\begin{abstract}
Large gradual solar energetic particle (SEP) events can pose a radiation risk to crewed spaceflight and a significant threat to near-Earth satellites however, the origin of the SEP seed particle population, how these particles are released, accelerated and transported into the heliosphere are not well understood. We analyse NOAA active region (AR) 12673, that was the source responsible for multiple large gradual SEP events during September 2017, and found that almost immediately after each significant eruptive event associated with SEPs an enhanced Si/S abundance ratio was measured by Wind, consistent with the previous work by Brooks et al. Hinode/EIS took data roughly 8~hours before the second SEP event on 2017 September 6 that allowed the regions of enhanced Si/S abundance ratio in the AR to be determined. We have shown that the AR contains plasma with elemental abundance values detected in situ by Wind. In particular, the plasma originates from the core of the AR, similar to Brooks et al., but in the moss (footpoints) associated with hot sigmoidal AR loops. The sigmoid, that contains highly fractionated plasma, erupts and propagates towards an Earth-connected magnetic null point, providing a direct channel for the highly fractionated plasma to escape and be detected in the near-Earth environment.

\end{abstract}

\keywords{Solar Energetic Particles (1491); Solar Active Region Magnetic Fields (1503); Solar Magnetic Reconnection (1504); Solar Flares (1496); Solar Coronal Mass Ejections (310); Solar Abundances (1474)}

\shorttitle{Sigmoid eruption drives gradual SEP event}
\shortauthors{Yardley et al.}

\section{Introduction} \label{sec:intro}

Solar energetic particles (SEPs), consisting of electrons, protons and heavy ions, are often accelerated to very high energies through magnetic reconnection processes associated with solar flares and shocks driven by coronal mass ejections (CMEs). Gradual SEP events \citep[for a review see][]{Desai2016}, where the particle flux remains high for several days, pose a radiation risk to crewed spaceflight and a threat to satellites. As the first SEPs arrive at Earth very quickly, within several minutes of an energetic event occurring at the Sun, forecasting SEP events in advance is therefore key to mitigating the risk posed by SEPs. This requires the identification of source regions prior to eruption and SEP generation. For a recent review of current SEP forecasting methods see \citet{Whitman2023}.

The primary sources of strong energetic events (flares, CMEs and SEPs) are magnetically complex active regions. To predict energetic events requires distinguishing eruptive vs non-eruptive active regions \citep{Georgoulis2024}. In particular, the prediction of SEP events requires not only forecasting whether an active region is likely to produce eruptive flares and CMEs, but also the magnetic field environment and magnetic connectivity of the source region (e.g. 
\citep[e.g.][]{Lario2013, Dresing2014, Yardley2022}, along with the location of the suprathermal seed particle population at the Sun.

Remote sensing and in-situ measurements are a powerful combination for understanding many aspects of eruptions and solar wind flows into the heliosphere. In the past, these measurements have been utilised to determine whether interplanetary CMEs exhibit signatures indicative of cool, dense plasma expected from filament eruptions \citep[e.g.][]{Lepri2010}. Elemental abundances, in particular, are extremely useful for tracing solar wind plasma detected in-situ to their coronal source regions, see e.g. \citet[][]{Brooks2015} and \citet[][]{Stansby2020} for studies connecting observations by the Advanced Composition Explorer (ACE; \citealt{Stone1998}) to their coronal origins with the EUV Imaging Spectrometer (EIS; \citealt{Culhane2007}) onboard Hinode \citep{Kosugi2007}, and most recently, \citep{Yardley2024} for an investigation using a combination of remote-sensing and in-situ instruments on board ESA/NASA's Solar Orbiter \citep{Muller2020}.

Easily ionised elements that have a First Ionisation Potential (FIP) $<$ 10~eV, known as low-FIP elements, are enhanced in the corona, SEPs and the solar wind compared to high-FIP elements (FIP $>$ 10 eV). This is due to the operation of the FIP effect, where the fractionation of ions and neutrals takes place in the upper chromosphere \citep{Laming2015}. A key spectroscopic technique has recently emerged based on Si and S that allows energetic particles detected at L1 to be traced back to their sources on the Sun \citep{Brooks2021a}. This is due to the behaviour of intermediate FIP elements, such as S, being dependent upon the magnetic field environment (open vs closed magnetic field). The behaviour of S suggests that it is more abundant in slow solar wind than the corona or during gradual SEP events.

\citet{Brooks2021a} used this SEP composition diagnostic to show that plasma confined by strong magnetic fields in the core, high-temperature loops rooted in the moss of an AR developed an enhanced Si/S abundance ratio with values as high as 4-5. Lower values were also measured adjacent to these core loops in the upflow regions at the edges of the AR.These values matched those detected in situ by the Wind spacecraft during the multiple particle events.

Therefore, the highly fractionated plasma can be released indirectly into the environment of the active region and the solar wind by interchange reconnection between the closed and open fields at the boundaries of active regions \citep{Brooks2021b}. Or the highly fractionated plasma can be released directly from the stronger core magnetic fields of the active region via flare reconnection. The plasma detected in situ is then a combination of plasma being released directly or indirectly from the AR core and/or boundary that is later accelerated by the CME-driven shock, often referred to as so called ``hybrid" events \citep[see e.g.][]{Vlahos2019}.


Here we focus on NOAA AR 12673 that produced three large gradual SEP events between 2017 September 4 and 10, the last of which was detected as a ground-level enhancement (GLE). AR 12673 is one of the most analysed active regions from solar cycle 24 \citep{Sun2017, Yang2017, Chertok2018, Cohen2018, Luhmann2018, Sharykin2018, Shen2018, Wang2018, Anfinogentov2019, Bruno2019, Romano2019, Moraitis2019, Yamasaki2021, Yardley2022, Joshi2023}. We show that an enhanced Si/S ratio was detected in situ by Wind during all three SEP events. We compare the in situ measurements of the Si/S abundance ratio taken by the Wind spacecraft during the second SEP event that occurred on 2017 September 6, to the ratio in the solar corona computed by applying spectroscopic techniques to Hinode/EIS data taken $\sim$8~hours before SEP occurrence. Using these techniques, we are able to establish the SEP source region of a second AR, in addition to the one studied by \citet{Brooks2021a}.

\section{Observations \& Methods} \label{sec:obs}

\begin{figure*}[ht!]
\plotone{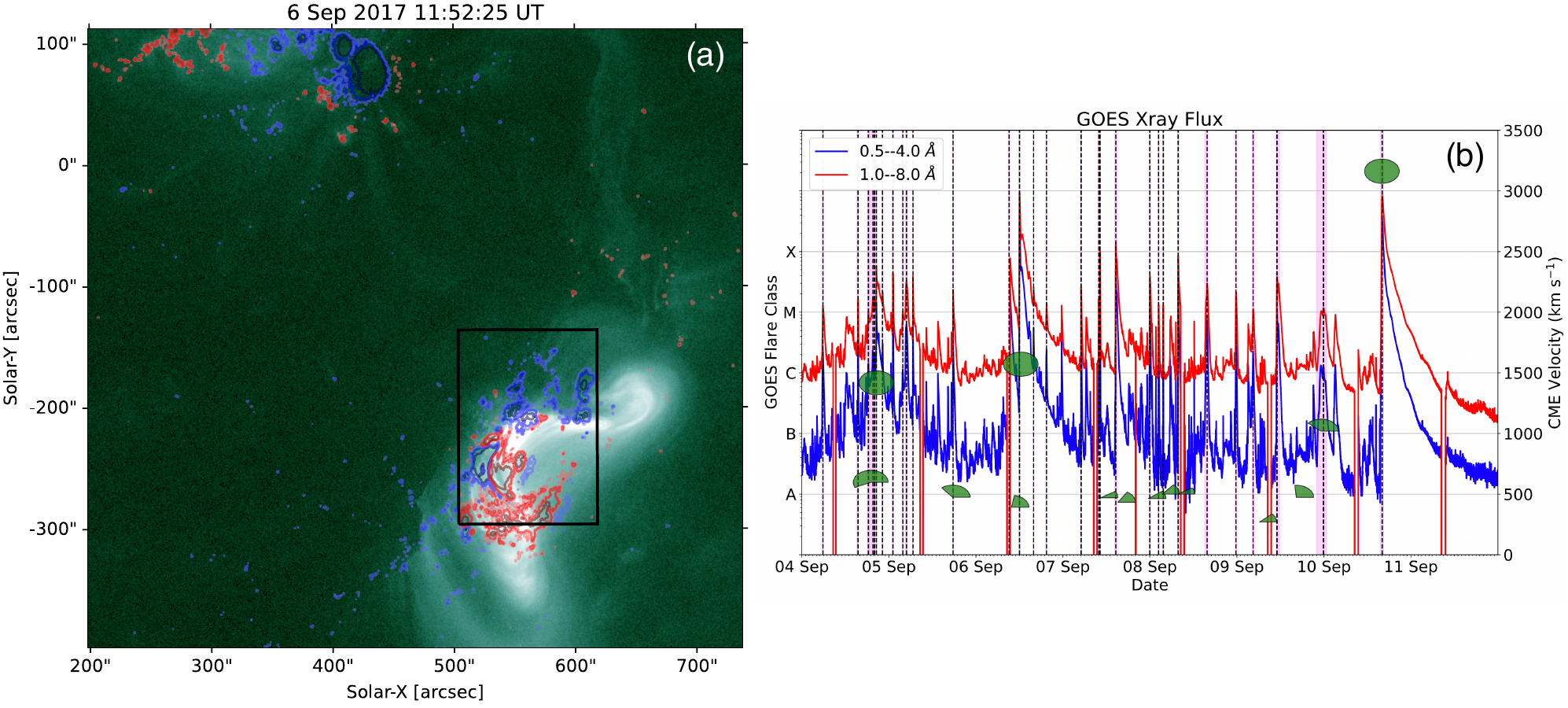}
\caption{(a) 94~\AA\ image of AR 12673 on 2017 September 6 at 11:49:28~UT just prior to the second SEP event taken by the Atmospheric Imaging Assembly (AIA; \citealt{Lemen2012}) onboard the Solar Dynamics Observatory (SDO; \citealt{Pesnell2012} The red (blue) contours correspond to the line-of-sight magnetic field taken by the Helioseismic and Magnetic Imager (HMI; \citealt{Scherrer2012}) on board SDO saturated at a level of 300~G (-300~G). The black rectangle represents the FOV of the Hinode/EIS observations taken $\sim$8~hours earlier. (b) The GOES-13 X-ray flux during the period 2017 September 4 to 12. The dashed lines represent the peak of each flare occurring during this period with the shaded regions indicating the duration of each flare. The green ovals show the occurrence and velocity of the CMEs that originated from AR 12673. The circle fraction represents the width of the CME taken from the SOHO LASCO CME catalogue (\url{https://cdaw.gsfc.nasa.gov/CME_list/}). \label{fig:fig1}}
\end{figure*}

\subsection{AR 12673 Evolution}
The source region (AR 12673) responsible for the SEP events of September 2017 was a highly complex region that remained visible for five Carrington rotations. The region first emerged in the southern hemisphere in July 2017 as AR 12665 with a Mount Wilson class of $\beta \gamma$. AR 12665 had a right-handed chirality,  which is unusual for the southern hemisphere. While Earth-facing, AR 12665 produced an SEP event on 14 July 2017 at 04:40~UT associated with an M2.4 class flare recorded by GOES at 01:07~UT and a CME first visible in LASCO/C2 at 01:25~UT. During the second rotation in August 2017, the region returned as AR 12670, and while the leading positive polarity of the region remained intact, the trailing negative polarity had decayed. There was no solar activity associated with this AR that was recorded during this rotation.

The positive leading spot was still apparent during the third rotation as AR 12673 (Figure~\ref{fig:fig1}~a). The magnetic polarities exhibited right-handed chirality, opposite to the first rotation, which is the expected trend in the southern hemisphere. A large amount of flux emerged into the region beginning on 2017 September 2. In fact, the region, which was given a $\beta \gamma \delta$ Mount Wilson classification, evolved to become highly complex (see Figure\ref{fig:fig1}~a). It has one of the fastest flux emergence rates \citep{Sun2017} along with the strongest transverse and coronal magnetic fields \citep{Wang2018, Anfinogentov2019} to date. The strong transverse fields were detected in a light bridge where \citet{Baker2020} also reported the presence of Inverse-FIP plasma, where low-FIP elements are depleted or high-FIP elements are enhanced, potentially caused by subsurface reconnection occurring between coalescing sunspots.

As expected, complex AR 12673 produced a large amount of significant (M and X class) flares and fast CMEs during its disk passage (Figure~\ref{fig:fig1}~b). Here we are particularly interested in the three SEP events, associated with some of the region's strongest flares and fastest, halo CMEs (Figure~\ref{fig:fig1}~b), that were produced on 2017 September 4, 6 and 10. There were many Hinode/EIS observations taken of AR 12673 during this time period. These observations included \ion{Si X} 258.375~\AA\ and \ion{S X} 264.223~\AA\ that can be used to compute abundance ratio maps, which can be compared to the Si/S abundance ratio measured in situ by the Wind spacecraft. In particular, we focus on the Hinode/EIS observations taken approximately 8~hrs before the second SEP event that occurred as a result of an X9.3 class flare (11:52~UT) and halo CME (12:24~UT) with a radial velocity of 2268~km~s$^{-1}$ \citep{Yardley2022}, due to its close proximity in time to the occurrence of the SEP event. At this time, a sigmoid (continuous S-shaped loops) have already formed in the north-western part of the AR. The FOV of the EIS observations can be seen in Figure~\ref{fig:fig1}~(a). Further analysis of the sigmoid and its eruption is discussed in \citet{Yan2018, Mitra2018}. After September 2017, this region reappeared for another two rotations until it eventually decayed into background quiet Sun magnetic flux.  

\begin{figure*}[ht!]
\plotone{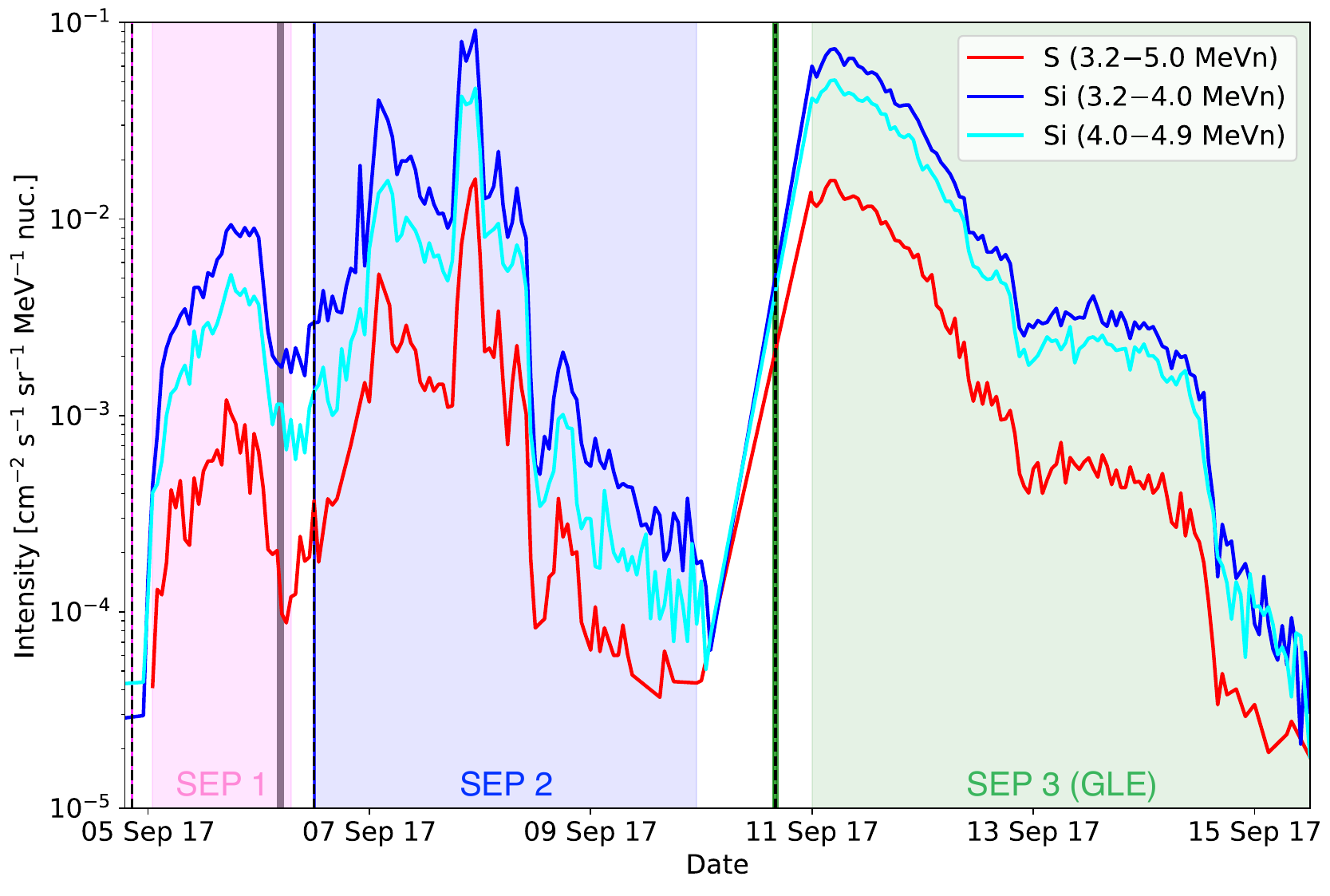}
\caption{The Si and S intensities measured by Wind EPACT/LEMT during the time period of 2017 September 5 to 15. The three different colours (red, dark and light blue) represent the three different energy channels. The pink, blue and green shaded areas highlight the three SEP events that were detected on 2017 September 4, 6 and 10. The grey shaded area shows the timing of the Hinode/EIS spectroscopic observations. The black dashed line shows the time of the X9.3 GOES class flare and the corresponding dark blue shaded area the flare duration. 
\label{fig:fig2}}
\end{figure*}

\subsection{Wind Particle Fluxes}

The hourly time series of Si intensities that were measured by the EPACT/LEMT (Energetic Particles: Acceleration, Composition and Transport/Low Energy Matrix Telescope; \citealt{vonRosenvinge1995}) onboard Wind \citep{Ogilvie1997} were downloaded from the NASA/Goddard Space Flight Center OMNIWeb Plus Browser. Two energy channels were used with low ranges, in particular the 3.20 - 4.00 MeV/n and 4.00 - 4.90 MeV/n channels. The S intensities are not publicly available from OMNIWeb, instead they were obtained from the EPACT team.

\subsection{Hinode/EIS elemental abundance measurements}

The Hinode/EIS data are obtained from DARTS (Data Archive and TRansition System), which is hosted at ISAS/JAXA (Institute of Astronautical Science/Japan Aerospace Exploration Agency). The SolarSoft IDL eis\_prep routine is applied to the EIS data to take into account pixels that are affected by dust, electrical charge, and cosmic ray hits. It also removes the CCD dark current pedestal. For the composition measurements we used an EIS raster scan with a FOV of 120'' $\times$ 160'' created with a 2'' slit, and an exposure time of 60~s.

The Si/S abundance maps are calculated using an established method (see \citet{Brooks2011, Brooks2015}). \ion{Si X} 258.375~\AA\ and \ion{S X} 264.223~\AA\ are subject to temperature and density effects and so a multitude of Fe lines are used to quantify these. The density is used to calculate the contribution functions under ionisation equilibrium for the Fe lines that span between 0.52 to 5.5~MK. The emission measure (EM) is then computed using the equation for spectral intensity by utilising the Markov-Chain Monte Carlo (MCMC) software from the PINTofALE package \citep{Kashyap1998, Kashyap2000}. The simulations are performed until the observed line intensities and temperature distribution are reproduced. The temperature distribution is then rescaled to match the Si X intensity. The Si/S abundance ratio is provided by taking the ratio of the predicted and observed S X intensities if we assume photospheric abundances. The accuracy of the EIS composition measurements is roughly 30~\% due to large calibration errors \citep{Brooks2015}.

\section{Results}

\subsection{In situ particle measurements}
The three gradual SEP events that occurred on 2017 September 4, 6, and 10 were detected in situ by the Wind spacecraft, the time periods of which have been shaded in pink, blue and green in Figure~\ref{fig:fig2}. The particle flux increases during the three SEP events were each preceded by a significant flare and CME originating from AR 12673. The second SEP event, which is the focus here, occurred after the X9.3 flare and fast halo CME that was described in the previous section (see Section~\ref{sec:obs}), the onset (duration) of which is highlighted by the black dashed line (orange shaded region) in Figure~\ref{fig:fig2}. The grey shaded region shows the times of the corresponding Hinode/EIS Si and S observations taken 8~hrs before the event.

We see a similar trend to Figure~4 in \citet{Brooks2021a} where the particle fluxes detected by Wind increase almost immediately after the occurrence of the flares/CMEs, suggesting that the particles were accelerated low down in the solar atmosphere by CME-generated shocks and flare reconnection processes. This could explain the production, injection and acceleration of the suprathermal seed particle population. The particle fluxes during the second and third SEP events are increased by at least one order of magnitude compared to the first, indicating an increasing accessibility to a larger particle seed population with time.

\begin{figure*}[ht!]
\plotone{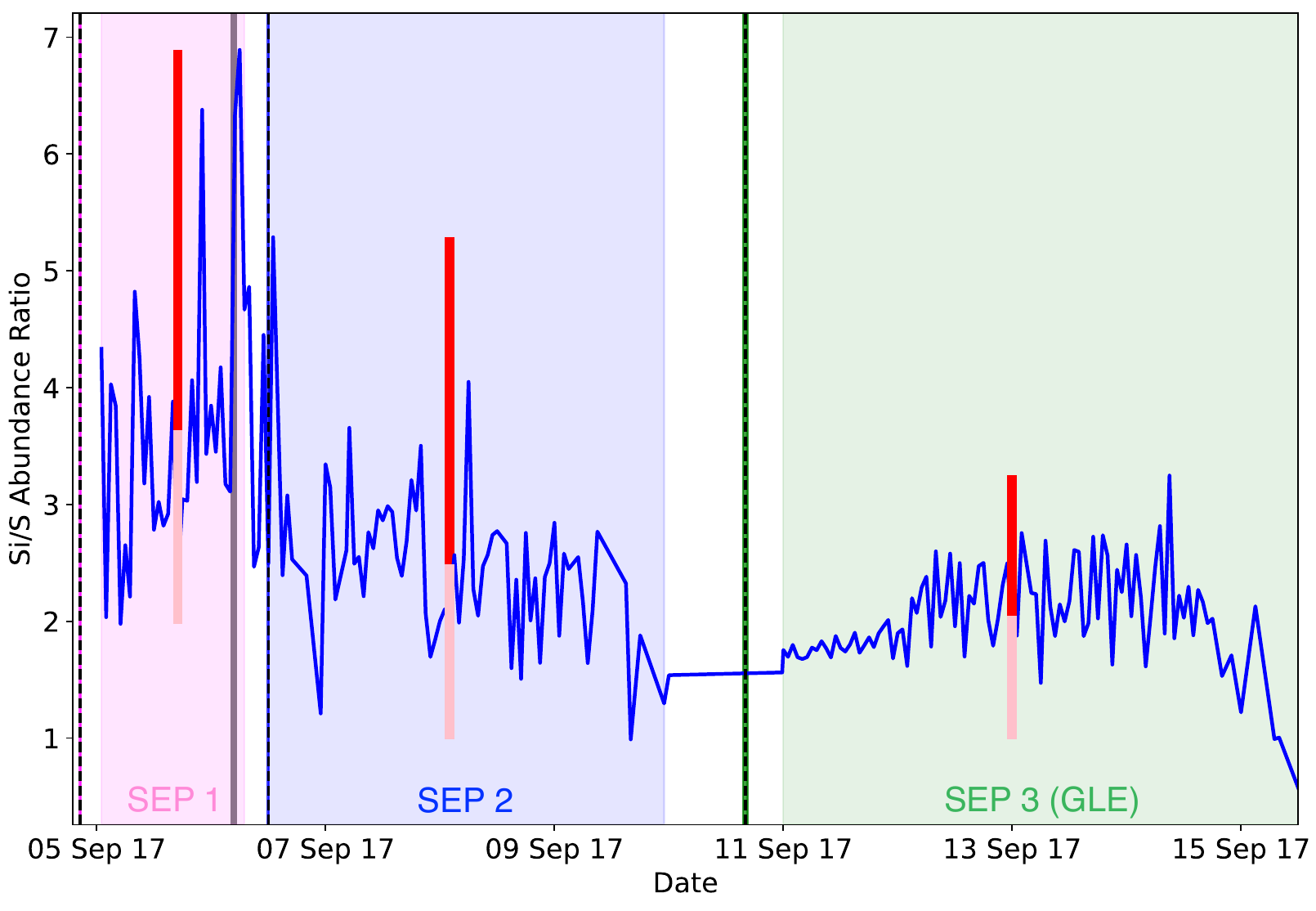}
\caption{The Si/S abundance ratio calculated using the EPACT/LEMT data for the time period in Figure~\ref{fig:fig2}. 
The ratio is calculated using an average of the two Si channels shown in Figure~\ref{fig:fig2} that have been normalised with respect to the photospheric abundance (2.34423 taken from \citealt{Grevesse2015}). The three red/pink colorbars show the minimum, mean and maximum Si/S abundance ratios for the three SEP events. The mean and maximum values for the second SEP event are used to plot the contours shown in panels (c) and (d) of Figure~\ref{fig:fig4}.
\label{fig:fig3}}
\end{figure*}

The corresponding time evolution of the Si/S abundance ratio measured by Wind is shown in Figure~\ref{fig:fig3}. The Si channels have been interpolated to match the hourly S data and values are only shown where the uncertainties in S are $<$~50~\%. There is a large range in values in the Si/S abundance ratio however, during the SEP events the ratio always remains above expected photospheric values. The time periods (pink, blue and green shaded regions in Figures~\ref{fig:fig2} and \ref{fig:fig3}) used to calculate the range of values of the Si/S ratio (pink and red contours in Figure~\ref{fig:fig3}) for each SEP event is given in Table~\ref{tab:tab1}. 

\begin{table*}[ht] \label{tab:tab1}
\centering
\begin{tabular}{cccccc}
\multicolumn{3}{c}{SEP Event}               & \multicolumn{3}{c}{Si/S Abundance Ratio} \\ 
No. & Start Time {[}UT{]} & End Time {[}UT{]} & Min          & Mean         & Max         \\ \hline
1   & 2017-09-05 01:00    & 2017-09-06 07:00  & 2.0          & 3.6          & 6.9         \\
2   & 2017-09-06 12:01    & 2017-09-09 23:01  & 1.0          & 2.5          & 5.3         \\
3   & 2017-09-11 00:01    & 2017-09-15 12:00  & 1.5          & 2.1          & 3.3        
\end{tabular}
\caption{The minimum, mean and maximum values of the Si/S abundance ratio calculated for each of the three SEP events (see shaded regions in Figure~\ref{fig:fig3}).}
\end{table*}

The event averages range from 2.1 to 3.6, which is consistently higher than reported in \citep{Brooks2021a} although, we have only analysed two active regions to date. The event averages for the second and third SEP events are closer to what we would expect as the typical SEP abundance is 2.0 whereas, for the first event it is almost double this value. Higher values of the Si/S abundance ratio are measured during certain periods for each event, with maximum values ranging between 3.3 and 6.9. Interestingly, both the mean and maximum Si/S abundance ratio values decrease with each consecutive SEP event. In the next section, we will use the Hinode/EIS observations taken prior to the second SEP event on 2017 September 6 to determine the locations in the AR where we find similar enhanced values of the Si/S abundance ratios.

\subsection{Hinode/EIS spectroscopic measurements}

\begin{figure*}[ht!]
\plotone{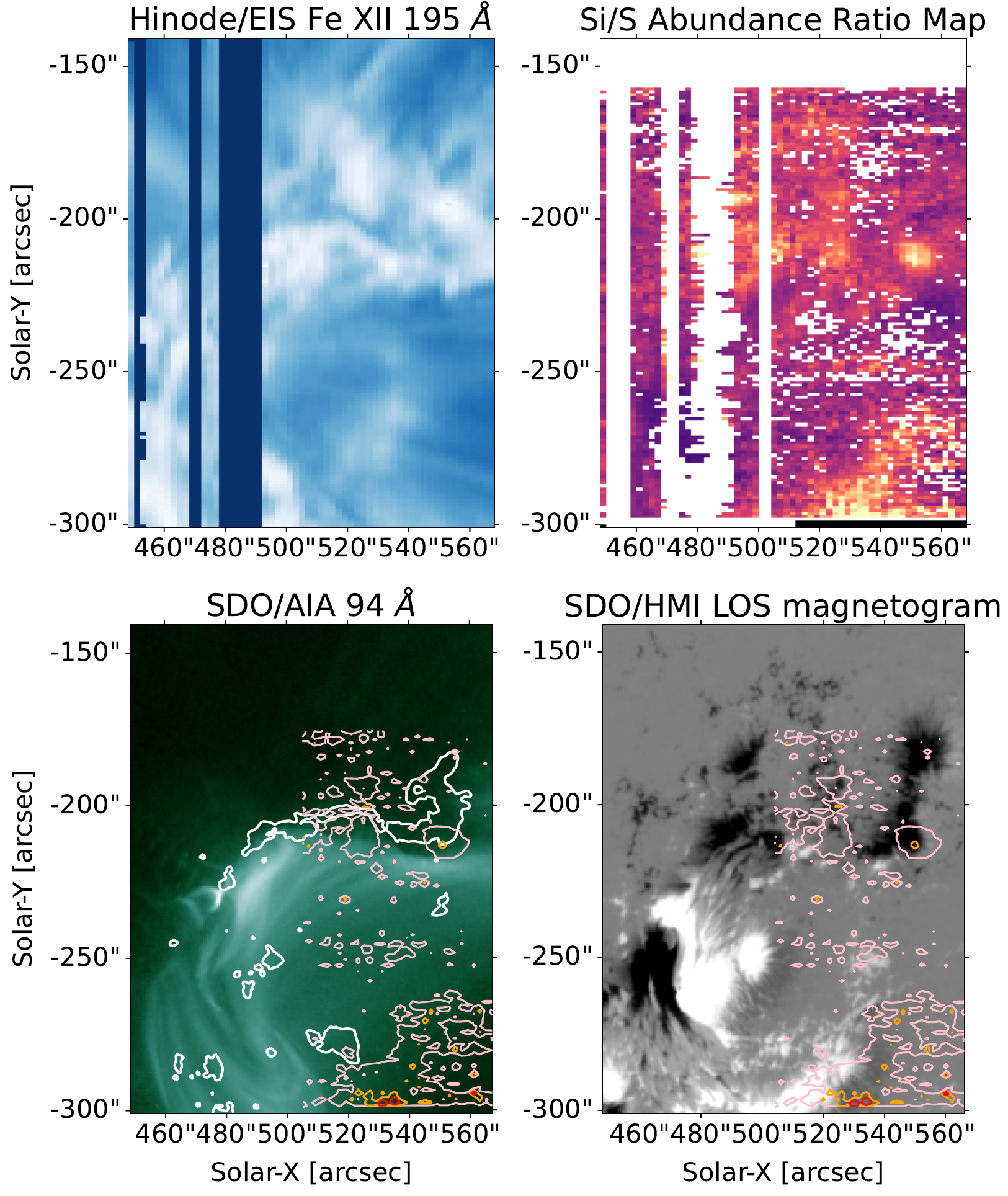}
\caption{Hinode/EIS and SDO observations of AR 12673 taken $\sim$8~hours prior (2017 September 6 at 04:42~UT) to the second SEP event. Panels~(a) and (b) show the Fe~XII 195~\AA\ Intensity and Si/S abundance ratio map produced from the Hinode/EIS observations. Panels~(c) and (d) show the corresponding  SDO/AIA 94~\AA\ and SDO/HMI LOS magnetic field observations. The pink and red contours show the locations of the minimum and maximum Si/S abundance ratio values during the second SEP event using the Hinode/EIS abundance ratio map, and the orange contours show the intermediate values. The white contours show the flare ribbons taken from AIA 1600~\AA\ data during the gradual phase of the flare (13:30~UT).   
\label{fig:fig4}}
\end{figure*}

In Figure~\ref{fig:fig4} we show the Hinode/EIS and corresponding SDO observations of AR 12673 with the same FOV (120''$\times$160'') as the EIS raster scan. The top row shows the EIS Fe XII intensity and the Si/S abundance ratio map computed using the EIS dataset. These maps can be noisy and so a mask was applied to remove any pixels with (i) a $\chi^{2}$ larger than the number of lines used (18), which implies that the computed intensities match the observed ones within the uncertainties, and (ii) with a calibration error larger than 25~\%. We then overlay contours taken from the masked Si/S abundance ratio map with values corresponding to a range in values taken from the the red/pink contours in Figure~\ref{fig:fig3} on the SDO data shown in the bottom row. Pink contours show the event average (2.5) whereas, red represent the maximum value (5.3), and finally orange show intermediate values (4.0).

From overlaying these contours on SDO/AIA it is evident that the AR contains plasma with an elemental abundance that matches that detected in situ by the Wind spacecraft. In particular, the plasma originates from high-temperature sigmoidal loops that are rooted in the bright moss regions (see Figure~\ref{fig:fig4}~c and d) in the north. The sigmoidal loops extend south but are rooted in the positive polarity adjacent to the other location of high Si/S abundance measurements. However, two flare ribbons (white contours in Figure~\ref{fig:fig4}~d), develop initially extending from the negative polarity in the north associated with the footpoints of the sigmoid and stretch across the core of the active region. Later on, during the gradual phase of the flare, the second flare ribbon reaches the positive active region polarity in the south whose core and edge are both associated with the high values of Si/S. These findings are similar to \citet{Brooks2021a} except the active region core loops are sigmoidal in this case. Strong coronal magnetic fields \citep{Wang2018, Anfinogentov2019} present in this AR could also explain why we detect areas of enhanced Si/S ratio in this AR, again similar to AR 11944.


\section{Summary and Discussion}

In this study, we compared the Si/S abundance ratio from particle events detected in situ by Wind to the corona by applying spectroscopic techniques to Hinode/EIS data. This analysis was carried out in order to determine whether the plasma with enhanced Si/S ratio detected in situ could be traced back to its source in the corona. We chose AR 12673, that was one of the most eruptive ARs in solar cycle 24, that produced multiple SEP events during September 2017. The SEP event that occurred on 2017 September 10 was associated with a ground-level enhancement (GLE). We discovered that AR 12673 showed similar SEP composition signatures to those detected in AR11944 \citep{Brooks2021a}.

We found that the particle fluxes measured by Wind increased dramatically following the occurrence of strong flares and fast, halo CMEs that were associated with the three SEP events (2017 September 4, 6 and 10) and remained high for many days. When calculating the Si/S abundance ratio for each event we found that the ratio was enhanced in all cases. In particular, the average enhancement that was calculated during each event was found to either be in agreement with or higher than the typical event averages for SEPs \citep{Reames2018}, with values ranging from 2.1 to 3.6. While, the maximum values of the Si/S abundance ratio ranged between 3.3 and 6.9, slightly lower and higher, respectively than that was found in the events analysed by \citet{Brooks2021a} (max values ranging from $\sim$4-5). Interestingly, while the particle fluxes measured during the second and third events were an order of magnitude higher than the first, the average and maximum values of the Si/S abundance ratio showed a decreasing trend. This could be due to the fact that as S behaves as an intermediate FIP element, more extreme events could energize the chromosphere at lower depths where S is more fractionated or that there is some relation to the direction of the open/closed magnetic field. The cause of the decreasing trend should be investigated further.

We then compared the Si/S abundance ratio measured by Wind during the second SEP event that occurred on the 2017 September 6 to elemental abundance measurements made in the AR $\sim$8~hrs in advance of this event. This is due to the availability of suitable EIS studies for this type of analysis. By overlaying contours taken from the Hinode/EIS Si/S abundance ratio map with values that matched the average and maximum Si/S abundance ratio measured by Wind we found that the plasma originates from multiple locations corresponding to the footpoints of high temperature sigmoidal loops that are rooted in the active region moss, the flare ribbons, and also concentrated at the core and boundary of the positive magnetic field polarity in the south.

\citet{Yardley2022} analysed the magnetic field configuration and Earth-connectivity of AR 12673 along with the propagation direction of the CMEs. They discovered that these factors were all important when it came to the acceleration of energetic particles and their arrival at Earth. It was reported that for this event on 2017 September 6 that the location of the Earth-connected field was to the north-east in AR 12674 where a magnetic null point was present (see Figure~4). The eruption that occurred propagated radially inwards and towards the Earth-connected null. The eruption drove reconnection at the null, which was observed in EUV (visible in Figure~\ref{fig:fig1}~a) and provided a direct channel for the highly fractionated plasma to either be released from the core of the active region via flaring or indirectly through reconnection, then shock-accelerated along Earth-connected field. A similar scenario was found for the CME associated with the first SEP event that took place on 2017 September 4, except that the Earth-connected field was located to the east of the AR, south of the null point, providing a more indirect channel for the particles to escape. A larger number of ARs should be analysed to determine this scenario is commonly observed.

\begin{acknowledgments}
The authors would like to thank Don Reames for providing the Wind Sulfur hourly data. S.L.Y. is grateful to the Science Technology and Facilities Council for the award of an Ernest Rutherford Fellowship (ST/X003787/1), and the work of D.H.B. was performed under contract to the Naval Research Laboratory and was funded by the NASA Hinode program.
Hinode is a Japanese mission developed and launched by ISAS/JAXA, collaborating with NAOJ as a domestic partner, and NASA and STFC (UK) as international partners. Scientific operation of Hinode is performed by the Hinode science team organized at ISAS/JAXA. This team mainly consists of scientists from institutes in the partner countries. Support for the postlaunch operation is provided by JAXA and NAOJ (Japan), STFC (UK), NASA, ESA, and NSC (Norway). We would like to thank the Wind and EPACT teams for providing the data publicly online, available from the NASA Space Physics Data Facility. This work made use of data from NASA's Solar Dynamics Observatory (SDO) and so we are grateful to the AIA and HMI science teams and also to Jhelioviewer \citep{Muller2017} for making it possible to view the data \citep{Muller2017}. This work also utilizes data from NSO/GONG, which is operated by AURA under a cooperative agreement with NSF and with additional financial support from NOAA, NASA, and USAF. This research made use of open source, community-developed SunPy \citep{sunpy2020} and spacepy \citep{spacepy2011, spacepy2022} Python packages. For the purpose of open access, the author has applied a ‘Creative Commons Attribution (CC BY) licence (where permitted by UKRI) to any Author Accepted Manuscript version arising. All of the data utilised in this manuscript are either available publicly online, from the instrument teams, or the authors.
\end{acknowledgments}

\vspace{5mm}
\facilities{SDO/AIA, SDO/HMI, Hinode/EIS, GOES, Wind}

\software{JHelioviewer \citep{Muller2017},  
          SunPy \citep{sunpy2020}, 
          spacepy \citep{spacepy2011, spacepy2022},
          }

\bibliography{ref}{}
\bibliographystyle{aasjournal}

\end{document}